\documentclass[traditabstract]{aa} 
\def\degr{\hbox{$^\circ$}}
\def\arcmin{\hbox{$^\prime$}}
\def\arcsec{\hbox{$^{\prime\prime}$}}

\def\micron{\hbox{$\mu$m}}

\usepackage{natbib}

\usepackage{graphicx}
\usepackage{txfonts}
\begin{document}
   \title{SDC13 infrared dark clouds: Longitudinally  collapsing filaments?\thanks{Based on observations carried out with the IRAM 30m Telescope. IRAM is supported by INSU/CNRS (France), MPG (Germany) and IGN (Spain).}}
   \author{N. Peretto\inst{1,2}, G. A. Fuller\inst{3},  Ph. Andr\'e\inst{2}, D. Arzoumanian\inst{4},  V. M. Rivilla\inst{5}, S. Bardeau\inst{6}, S. Duarte Puertas\inst{7}, J. P. Guzman Fernandez\inst{7}, C. Lenfestey\inst{3}, G.-X. Li\inst{8}, F. A. Olguin\inst{9,10},  B. R. R\"ock\inst{11,12}, H. de Villiers\inst{13}, J. Williams\inst{3} 
                     }
             \institute{School of Physics \& Astronomy, Cardiff University, Queens Buildings, The parade, Cardiff CF24 3AA, UK \\
              \email{Nicolas.Peretto@astro.cf.ac.uk}
               \and Laboratoire AIM, CEA/DSM-CNRS-Universt\'e Paris Diderot, IRFU/Service d'Astrophysique, C.E. Saclay, France
               \and Jodrell Bank Centre for Astrophysics, School of Physics and Astronomy, University of Manchester, Manchester, M13 9PL, UK
               \and IAS, CNRS (UMR 8617), Universit\'e Paris-Sud, B\^atiment 121, 91400 Orsay, France
                  \and Centro de Astrobiolog\'ia (CSIC-INTA), Ctra. de Torrej\'on-Ajalvir, km. 4, E-28850 Torrej\'on de Ardoz, Madrid, Spain
                \and Institut de Radioastronomie Millim\'etrique, 300 Rue de la piscine, F-38406 Saint Martin d'H\`eres, France
                  \and Universidad de Granada, 18071 Granada, Spain
               \and Max-Planck Institut f\"ur Radioastronomie, Auf dem H\"ugel, 69, 53121 Bonn, Germany
                \and School of Physics and Astronomy, University of Leeds, Leeds LS2 9JT, UK
               \and Departamento de Astronom\'ia, Universidad de Chile, Casilla 36-D, Santiago, Chile       
               	\and Instituto de Astrofisica de Canarias, E-38200 La Laguna, Tenerife, Spain 
		\and Universidad de la Laguna, Dept. Astrofisica, E-38206 La Laguna, Tenerife, Spain 
		  \and Centre for Astrophysics Research, University of Hertfordshire, College Lane, Hatfield, Herts, AL10 9AB, UK
               }

   \date{Received; accepted}

 
  \abstract
 {Formation of stars is now believed to be tightly linked to the dynamical evolution of interstellar filaments in which they form. In this paper we analyze the density structure and kinematics of a small network of infrared dark filaments, SDC13, observed in both dust continuum and molecular line emission with the IRAM 30m telescope. These observations reveal the presence of 18 compact sources amongst which the two most massive, MM1 and MM2, are located at the intersection point of the parsec-long filaments. The dense gas velocity and velocity dispersion observed along these filaments show smooth, strongly correlated, gradients. We discuss the origin of the SDC13 velocity field in the context of filament longitudinal collapse. We show that the collapse timescale of the SDC13 filaments (from 1~Myr to 4~Myr depending on the model parameters) is consistent with the presence of Class I sources in them,  and argue that, on top of bringing more material to the centre of the system, collapse could generate additional kinematic support against local fragmentation, helping the formation of starless super-Jeans cores. 
 }

\authorrunning{Peretto et al.}
   \maketitle
   
%

\section{Introduction}

In recent years, interstellar filaments have received special attention. The far-infrared {\it Herschel} space observatory revealed the ubiquity of filaments  in both quiescent and active star-forming clouds. The detailed analysis  of large samples of filaments identified with {\it Herschel} suggest that they represent a key stage towards the formation of low-mass prestellar cores \citep{arzoumanian2011}. In the picture proposed by \citet{andre2010}, these cores form out of filaments which have reached the thermal critical mass-per-unit-length, $M_{line,crit}^{th}=2c_s^2/G$ \citep{ostriker1964}, above which filaments become gravitationally unstable. Filaments with $M_{line} >M_{line,crit}^{th}$ are called supercritical filaments.
While this scenario might apply to the bulk of the low-mass prestellar cores, it can hardly  account for the formation of super-Jeans cores \citep[e.g.][]{sadavoy2010} whose mass is several times larger than the local thermal Jeans mass $M_{J}^{th}$ ($\sim1$~M$_{\odot}$ for critical 10~K filaments). Additional support (magnetic, kinematic) and/or significant subsequent core accretion are necessary to explain the formation of cores with masses $M_{core} >>M_{J}^{th}$.

Several high-resolution studies of  filamentary cloud kinematics have been performed in the past three years towards low-mass, nearby, star forming regions \citep[e.g.][Arzoumanian et al. 2013]{duartecabral2010, hacar2011, kirk2013}. These studies demonstrate the importance of cloud kinematics in the context of filament evolution and core formation. For instance, \citet{kirk2013} showed that gas filamentary accretion towards the central cluster of young stellar objects (YSOs) in Serpens South could sustain the star formation rate observed in this cloud, providing enough material to constantly form new generations of YSOs. But, it is not clear if these inflows have any impact at all in determining the mass of individual low-mass cores. 
On the high-mass side, infrared dark clouds (IRDCs) have been privileged targets \citep[e.g.][]{miettinen2012, ragan2012, henshaw2013,busquet2013, peretto2013}. These sources are typically located at a distance of $\sim4$~kpc from the sun \citep{peretto2010a}, making the analysis of filament kinematics more difficult. Thanks to the high sensitivity and resolution of ALMA, \citet{peretto2013} showed that  the global collapse of the SDC335.579-0.292 IRDC is responsible for the formation of an early O-type star progenitor ($M\simeq545$~M$_{\odot}$ in 0.05~pc) sitting at the  cloud centre. 

In this paper, we present results on an IRDC (hereafter called SDC13), composed of three Spitzer Dark Clouds from the Peretto \& Fuller (2009) catalogue (SDC13.174-0.07,SDC13.158-0.073, SDC13.194-0.073). The near kinematical distance of SDC13 is $d=3.6 (\pm 0.4)$~kpc using the \citet{reid2009} model. From 8\micron\ extinction we estimate a total gas mass of $\sim600$~M$_{\odot}$ above $N_{H_2}>10^{22}$~cm$^{-2}$.  As any IRDC, SDC13 shows little star formation activity, except in the central region where extended infrared emission is observed (see Fig.~1 from \citealt{peretto2010a}). The high extinction contrast, the rather simple geometry, and the large aspect ratios of the SDC13 filaments are remarkable, and provide a great opportunity to better understand the impact of filament kinematics on the formation of cores in an intermediate mass regime.

 \begin{figure}
 \hspace{0.8cm}
   \includegraphics[width=7.cm]{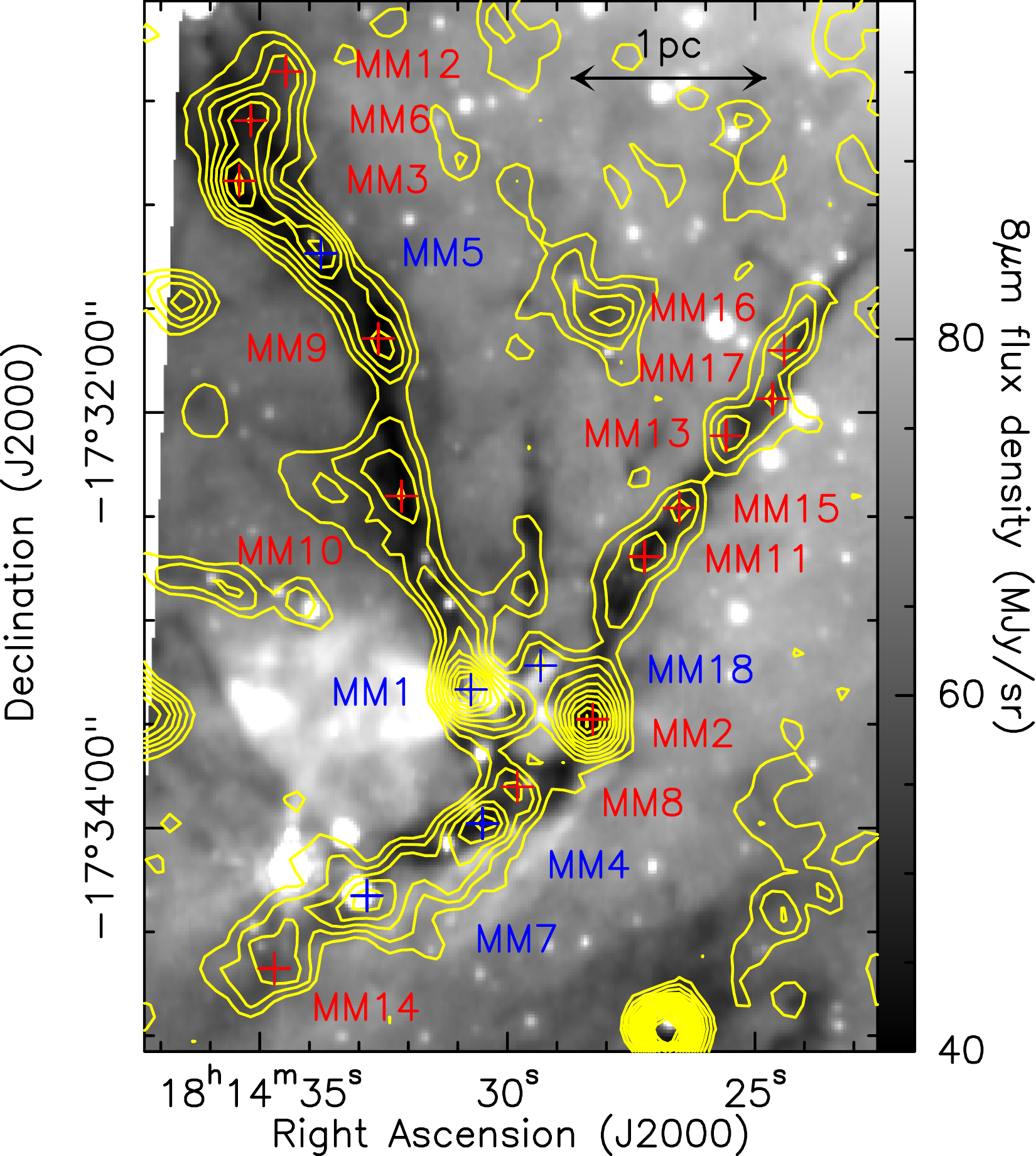}
      \caption{Spitzer 8\micron\ image of SDC13 in grey scale on top of which we overlaid the IRAM 30m MAMBO 1.2mm dust continuum contours (from 3~mJy/beam to 88~mJy/beam in step of 5~mJy/beam). The positions of the identified 1.2mm compact sources within the SDC13 filaments are marked as crosses, red for starless sources and blue for protostellar sources. 
              }
         \label{continuum}
   \end{figure}

\section{Millimeter observations}
\subsection{Dust continuum data}
In December 2009 we observed SDC13 at the IRAM 30m telescope using the MAMBO bolometer array at 1.2mm. We performed on-the-fly mapping of a $5\arcmin\times5\arcmin$ region. The sky opacity at 225~GHz was measured to be between 0.08 and 0.26 depending on the scan. Pointing accuracy was better than 2\arcsec\ and calibration on Uranus was better than 10\%. We used MOPSIC to reduce these MAMBO data, obtaining a final rms noise level on the reduced image of $\sim1$~mJy/11\arcsec-beam.


We also make use of publicly available {\it Spitzer}  GLIMPSE \citep{churchwell2009} and 24\micron\ MIPSGAL \citep{carey2009} data.

\subsection{Molecular line data}

In September 2011, during the IRAM 30m summer school, we observed SDC13 with the IRAM 30m telescope using the EMIR heterodyne receiver. We performed 31 single pointing observations along the SDC13 filaments in N$_2$H$^+$(1-0), at 93.2~GHz (i.e. 27\arcsec\ beam). The off positions, taken 5\arcmin\ away from SDC13, were checked to ensure no emission was present there. The pointing was better than 5\arcsec\ and the sky opacity  was varying between 0.3 to 2 at 225~GHz. We used the FTS spectrometer with a 50~KHz spectral resolution. The data have been reduced in CLASS\footnote{http://www.iram.fr/IRAMFR/GILDAS}. The resulting spectra have a rms noise varying from 0.05 to 0.1 K in a 0.16~km/s channel width.

  \begin{figure}
 \hspace{-0.3cm}
   \includegraphics[width=9.1cm]{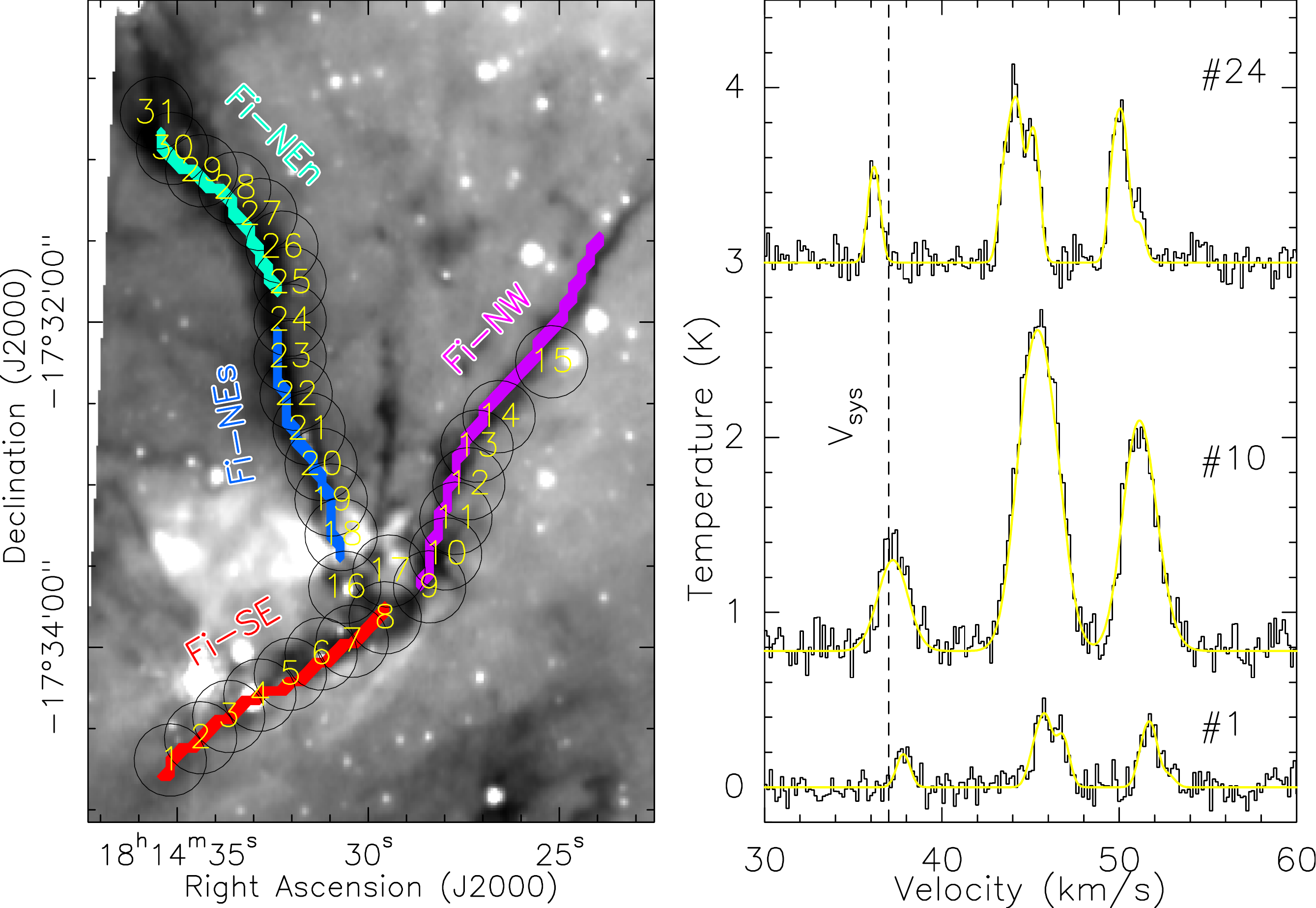}
      \caption{(left): Spitzer 8\micron\ image of SDC13 in grey scale on top of which we marked the 31 single-pointing positions we observed in N$_2$H$^+$(1-0), and the skeletons of each  filament (thick solid lines). The size of the black circles represents the 30m beam size at 3.2mm. (right): Examples of three  N$_2$H$^+$(1-0) spectra (displayed in T$_a^*$ temperature scale) observed at three different positions (1, 10, and 24). Their corresponding hyperfine structure fits are displayed as yellow solid lines. The remaining spectra are displayed in online Fig.~\ref{allspec}. The vertical dashed line corresponds to the systemic velocity of the cloud (i.e. $V_{sys}=37.0$~km/s) as measured from the isolated component of the hyperfine structure.}
         \label{spectra}
   \end{figure}

\section{Analysis}

\subsection{Mass partition in SDC13}

The MAMBO observations of SDC13 are presented in Fig.~\ref{continuum}. We can see that the 1.2mm dust continuum emission towards SDC13 matches the mid-infrared extinction very well. In the remainder of this paper we focus only on the emission towards these dark filaments and ignore any other significant emission peaks since we do not have kinematical information for those. 

We segmented the emission on this  dust continuum map into four filaments: Fi-SE, Fi-NW, Fi-NEs, Fi-NEn. The skeletons of these filaments  (Fig.~\ref{spectra}-left) have been obtained by fitting polynomial functions to the MAMBO emission peaks along them. The splitting of Fi-NE into two parts is due to its dynamical structure (cf Sect.~3.2). These filaments are barely resolved in our MAMBO map and are responsible for most of the 1.2mm flux  in the region. Although ground based dust continuum observations filter out all emission whose spatial scale is larger than the size of the bolometer array ($\sim2\arcmin$ for MAMBO), inspection of {\it Herschel} Hi-GAL observations \citep{molinari2010} shows that the SDC13 mass is indeed mostly concentrated within the filaments. The observed properties of the filaments (as measured within the 3~mJy/beam contour) are presented in Table 1\footnote{The sizes are estimated in the same way as for the IRDC catalogue of \citet[see Appendix A]{peretto2009}, meaning that we calculate the matrix of moment of inertia of the pixels within the  3~mJy/beam contour for each filament. Then we diagonalise the matrix and get two values corresponding to the sizes along the major and minor axes of the filaments. The radius R$_{eff}$ corresponds to the radius of the disc having the same area as the filament. Note that the  3~mJy/beam contour encompasses all filaments, we therefore artificially divided the MAMBO emission at the filaments junction.}.

Using the source identification scheme presented in Appendix A of \citet{peretto2009} we further identified 18 compact sources sitting within the filaments. 
The observed properties of these sources are given in Table~2. Note that a number of sources have sizes that are close to, or even smaller than, the beam size (all sources with dashes in Table 2 are in this situation). These sources have significant emission peaks ($>3\sigma$) but because they are  blended with nearby sources their measured sizes appear smaller than the beam, and their properties are very uncertain. All these sources are confirmed with higher resolution extinction map and SABOCA data (not shown here). The flux uncertainties in Table 2 reflect the difficulty to estimate the separation between the sources and the underlying filament. It is calculated by taking the average of the clipping and the bijection schemes of the flux estimates \citep{rosolowsky2008}. We further characterize the compact sources by checking their association with {\it Spitzer} 24\micron\ point-like sources, and their fragmentation as seen in the 8\micron\ extinction map (see Fig.~1 from Peretto \& Fuller 2010). We find that most of the sources are starless (13 out of 18) and that  only MM10 appears to be sub-fragmented in extinction. The results of this analysis are also quoted in Table 2.

Assuming that the dust emission is optically thin at 1.2mm, the measured fluxes provide a direct measurement  of the source mass. In Tables 3 and 4 we give the masses of all structures estimated using   a specific dust opacity at 1.2mm $\kappa_{1.2mm}=0.005$~cm$^2$\,g$^{-1}$, and dust temperatures of 12~K for filaments and starless sources, and 15~K for protostellar ones \citep{peretto2010b}. We further provide their volume densities, and the mass per unit length of filaments. All uncertainties result from the propagation of the flux uncertainties. They do not take into account systematic uncertainties on the dust temperature/opacity which account for an additional factor of $\le2$ uncertainty on all calculated properties.


   \begin{figure}
   \centering
   \includegraphics[width=7.2cm]{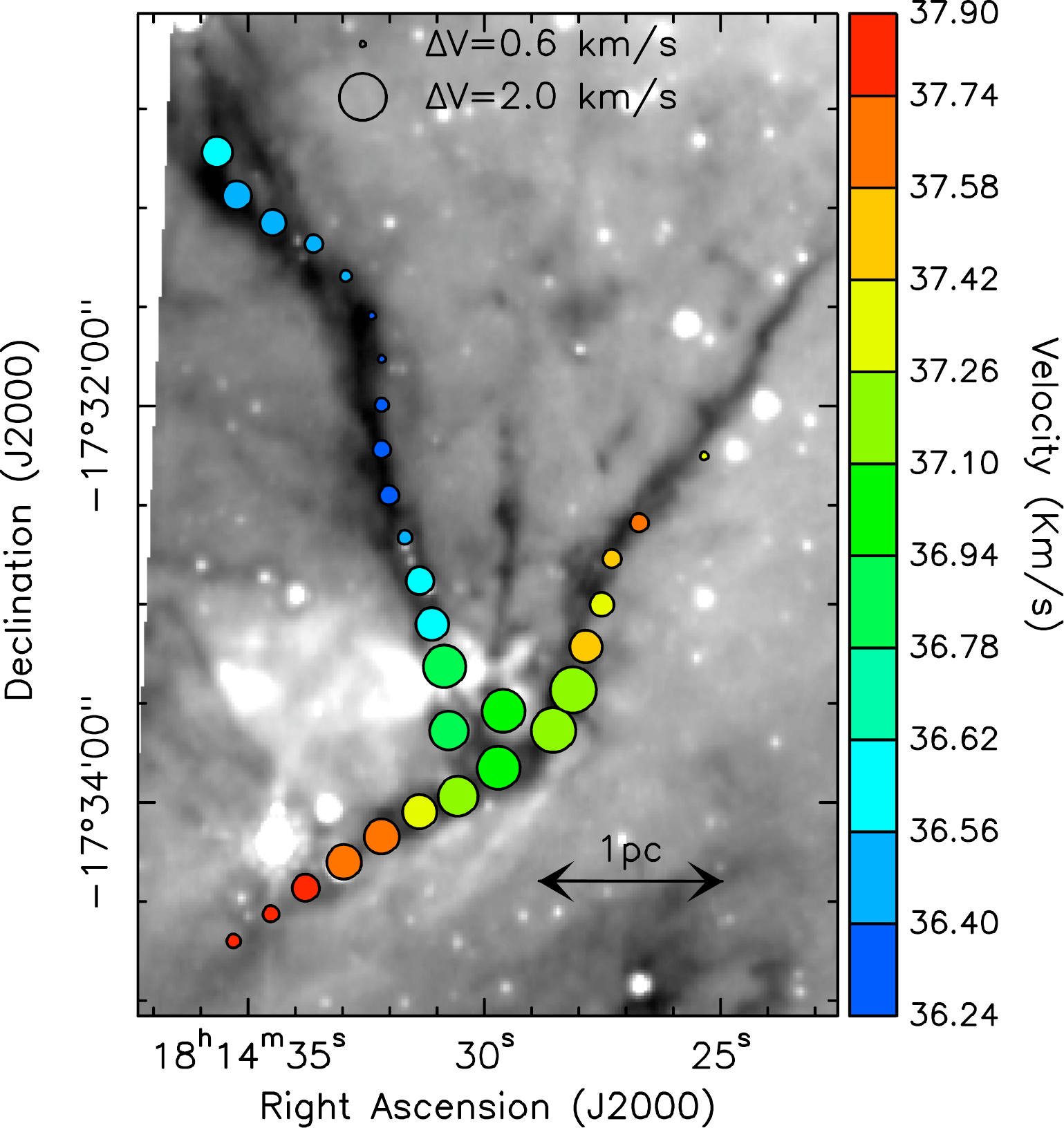}
      \caption{Spitzer 8\micron\ image of SDC13 in grey scale on top of which we symbolized the results of the N$_2$H$^+$(1-0) HFS fitting as circles. The colour of the symbols code the gas velocity while {\bf their} sizes code the gas velocity dispersion (FWHM). 
              }
         \label{velocity}
   \end{figure}

\subsection{Dense gas projected velocity field}

The N$_2$H$^+$(1-0) spectra observed along the SDC13 filaments are displayed in Fig.~\ref{spectra} and online Fig.~\ref{allspec}.  All spectra show strong detections ($S/N\ge4$) for all hyperfine line components, and symmetric line profiles. Based on these observations we estimate the systemic velocity of SDC13 at +37~km/s (near position 8 in Fig.~\ref{spectra}). Note that from position 18 to 24 and 28 to 31  we observe a secondary velocity component separated by $+17$~km/s. In Fig.~\ref{spectra} it can be identified as a small bump at $V\simeq+54$~km/s on spectrum \#24. This higher velocity component most likely belongs to another cloud on the eastern side of SDC13 which overlaps, in projection, with the Fi-NE filament. For the remainder of the paper we decided to ignore this higher velocity component.

 We used the hyperfine structure line fitting routine of the CLASS software  to derive line parameters such as the centroid gas velocity and gas velocity dispersion (see Fig.~\ref{spectra}). We represent the results from this fitting procedure on Fig.~\ref{velocity}. The colour and size of the circular symbols code the line-of-sight velocity and linewidth, respectively, of the gas as measured in N$_2$H$^+$(1-0). The exact values are given in Table 5 for each position. 
Figure~\ref{velocity} shows that both the dense gas velocity and velocity dispersion vary smoothly along each filament, from one end to the other. The two quantities are strongly correlated. Along Fi-NE, we can actually see that the smooth variation of the velocity reaches a minimum (around position 25) to start increasing again towards position 31. We believe that this shows that the northern and southern parts of Fi-NE are dynamically distinct. This is the reason why we decided to subdivide this filament into two,  Fi-NEn and Fi-NEs.
The Fi-NE filaments are blue-shifted with respect to the systemic velocity of SDC13 while the Fi-NW and Fi-SE filaments are redshifted. The filaments  spatially and dynamically converge near position 8 at the systemic velocity of the system, i.e +37 km/s. At this same position, the linewidth reaches a maximum of $\sim2$~km/s, while at the filament ends the linewidth is the narrowest, i.e. $\sim0.7$~km/s.

\section{Discussion}

\subsection{Timescales and  filament stability}

The {\it Spitzer} protostellar objects we identify in SDC13 filaments are typically embedded Class I (or young Class II) objects (see Appendix B). The age  $\tau_{CI}$ of such objects, from prestellar core stage to Class I protostellar stage, is $1~\rm{Myr}\leq\tau_{CI}\leq2$~Myr \citep[e.g.][]{kirk2005,evans2009}, consequently setting a lower limit of 1~Myr for the age of the filaments themselves. Their widths\footnote{Note that these widths are compatible with the standard 0.1~pc FWHM width observed towards nearby interstellar filaments \citep{arzoumanian2011} since the SDC13 measurements we provide here extend beyond the FWHM of the filament profiles. } as measured on the MAMBO map are  $\sim 0.3$~pc. Taking a 3D velocity dispersion $\sigma_{3D}=\sqrt{3}\sigma_{tot}=1.0(\pm0.3)$~km/s we obtain a crossing time $t_{cross}=0.3(\pm0.1)$~Myr.  Given their lower age limit of 1~Myr, the SDC13 filaments must therefore either be gravitationally bound or confined by ram/magnetic pressure. The thermal  critical mass-per-unit-length  for a 12~K isothermal filament is  $M_{line,crit}^{th}=19$~M$_{\odot}$/pc, the observed values for the SDC13 filaments are higher by a factor of 4 to 8 (cf Table 3), meaning that they are thermally supercritical. However, if we consider non-thermal motions as an extra support against gravity the picture slightly changes. We define the effective critical mass-per-unit-length as $M^{eff}_{line,crit}=2\overline{\sigma_{tot}}^2/G$, where $\overline{\sigma_{tot}}$ is the average total velocity dispersion along the filaments. Taking  $\overline{\sigma_{tot}}=0.6$~km/s (cf Table~5)  we obtain $1\leq \frac{M^{eff}_{line,crit}}{M_{line}} \leq 2$ for all filaments, making them just bound. This ratio is probably slightly overestimated since the contribution from the longitudinal collapse of filaments (see following sections) probably contribute to the observed line widths in the form of non-supportive bulk motions.  The SDC13 filaments match the relationship between $M_{line}$ and $M^{eff}_{line,crit}$ found by \citet{arzoumanian2013} for thermally supercritical filaments.


  \begin{figure}
   \centering
   \includegraphics[width=4.4cm]{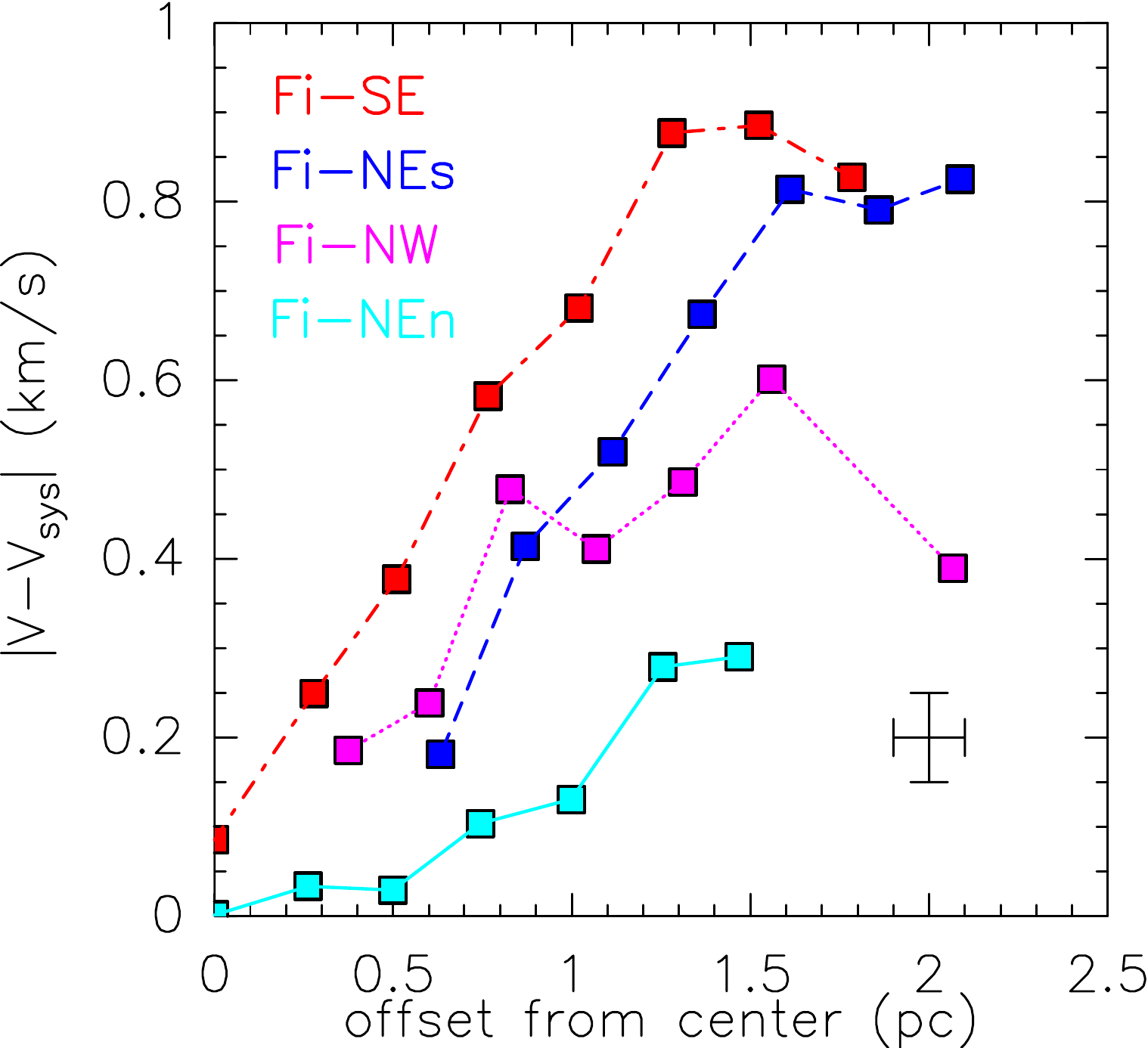}
      \includegraphics[width=4.4cm]{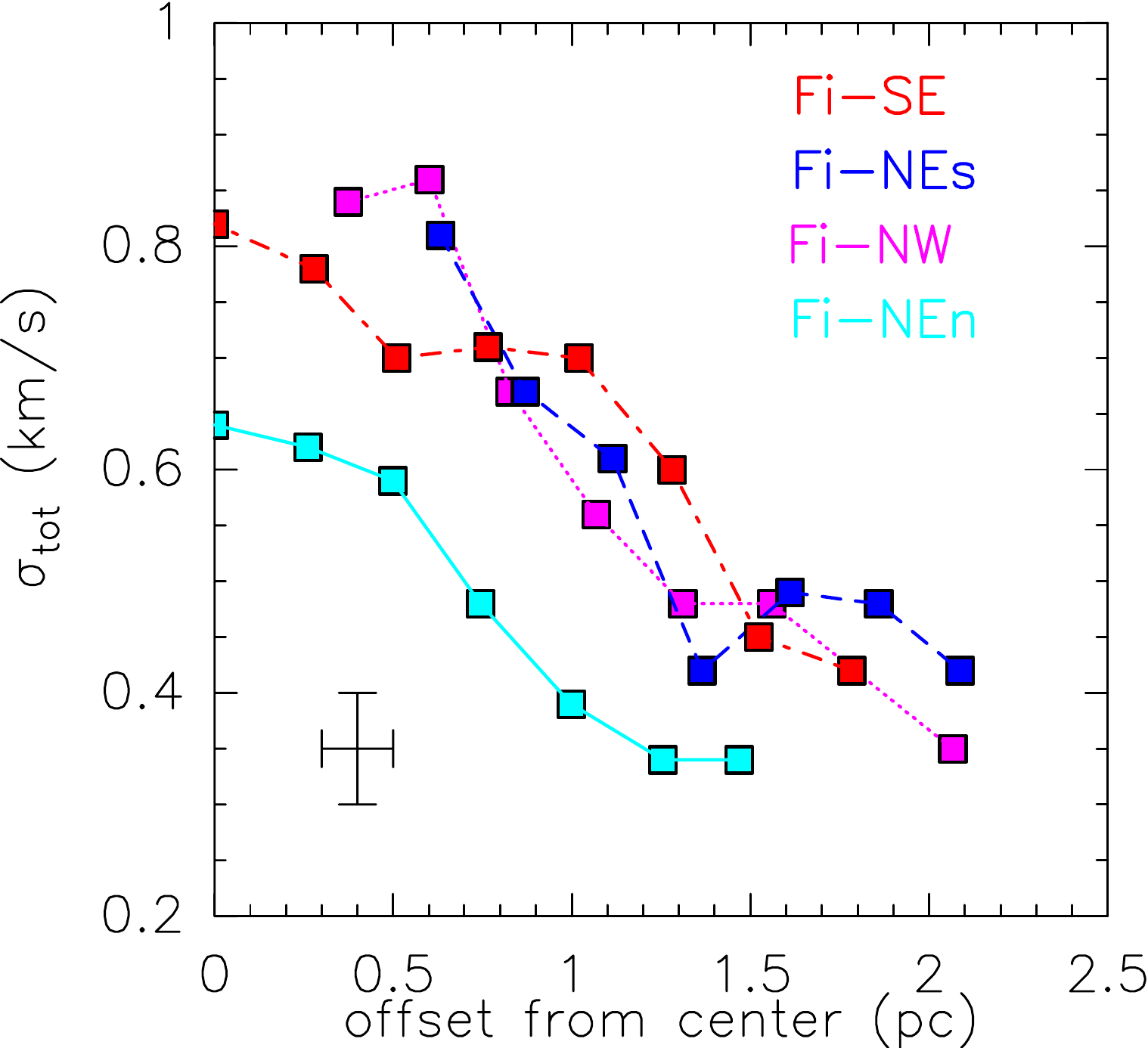}
      \caption{(left): Line-of-sight velocity profile of the dense gas within each filament as a function of position form the well centres, i.e. position 8 for Fi-NW (purple dotted line) , Fi-SE  (red dashed-dotted line), Fi-NEs (blue dashed line), and position 31 for Fi-NEn (cyan solid line). (right): Same as panel (left) but for the total (thermal+non-thermal) dense gas velocity dispersion. }
      \label{profiles}
      \end{figure}

\subsection{Dynamical evolution scenarios}

The smooth velocity structure observed along the SDC13 filaments shows that these filaments are dynamically coherent. The number of physical processes leading to a well organized parsec-scale velocity pattern, as observed in SDC13, is rather limited. Collapse, expansion, rotation, soft filament collision, and wind driven acceleration are the five main possibilities. Rotation would imply that the SDC13 filaments rotate about an axis going through the filament junction. The resulting geometry is unrealistic in the context of interstellar cloud dynamics, differential rotation would tear apart each filament in a crossing time $t_{cross}\simeq 0.3$~Myr (cf Sec.~4.1). No expanding source of energy (such as HII region or cluster of outflows\footnote{We observed few core positions in SiO(2-1), well known tracer of protostellar outflows, but did not get any wing emission.}) is located at the centre of SDC13, rendering the expansion scenario unlikely, while colliding filaments cannot account for the velocity pattern observed along Fi-NEn. Finally, wind-driven acceleration powered  by a nearby stellar cluster is a credible scenario. In particular, as we can see in Fig.~\ref{continuum} a mid-infrared nebulosity, along with a number of infrared compact sources, are observed east of MM1, probably interacting with it, and impacting its velocity through expanding winds. However, it probably does not impact the dynamics of the entire networks of filaments. Similarly to the colliding flow scenario, the wind-driven scenario can hardly explain the smooth velocity field observed in these filaments, and in particular, the one along Fi-NEn.
   
 We therefore consider that longitudinal collapse is the best option to explain the velocity structure observed in SDC13. Gravity can naturally account for a velocity gradient along the filaments as the gas collapses towards the centre of the system. In this context, one can argue for the presence of a primary gravitational potential well centre near position 8 towards which all the gas is collapsing, and a secondary one near position 31 towards which only Fi-NEn is infalling. In this collapse scenario, the redshifted filaments, i.e. Fi-NW and Fi-SE, are in the foreground with respect to the primary centre of the system, while the blueshifted ones, Fi-NEs and Fi-NEn, are in the background. The dense gas velocity observed along the filaments is therefore interpreted as a projected infall velocity profile. Figure~\ref{profiles} displays these profiles for all four filaments. We can see that for each filament the velocity is a linear function  of the distance to the centre (gradients are given in Table 3) up to $\sim1.5$~pc and then flattens out. This is expected in case of homologous free-fall collapse of filament having centrally concentrated density profiles along their crest (see Fig.~6 from \citealt{pon2011}, Fig.~5 from Peretto et al. 2007, and Fig.~1 from \citealt{myers2005}).

\begin{figure}
   \centering   
    \includegraphics[width=7.cm]{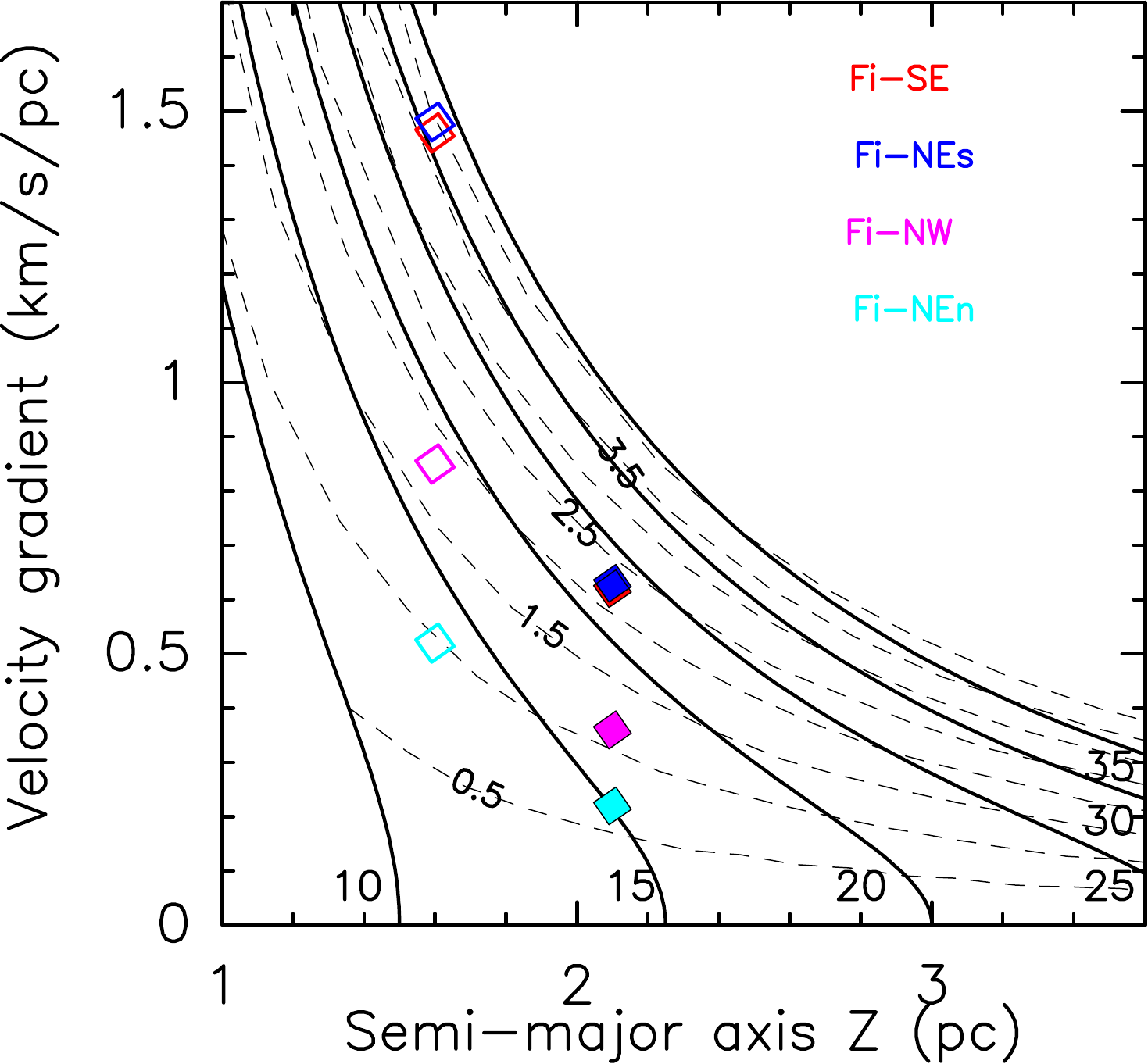}
      \caption{Time evolution of free-falling filaments in the $Z-\nabla V$ parameter space.  All filaments have the same initial density $n_0=4\times10^4$~cm$^{-3}$. Each solid line represents the evolution of a filament with a different initial aspect ratio $A_0$ indicated at the bottom end. Dashed lines represent isochrones from 0.5~Myr to 4~Myr separated by 0.5~Myr. The solid and empty squares mark the positions of the four filaments,  for projection angles of 45\degr\ and 67\degr, respectively (cf text and Appendix A).}
         \label{model}
   \end{figure}

\begin{figure}
   \centering
   \includegraphics[width=9cm]{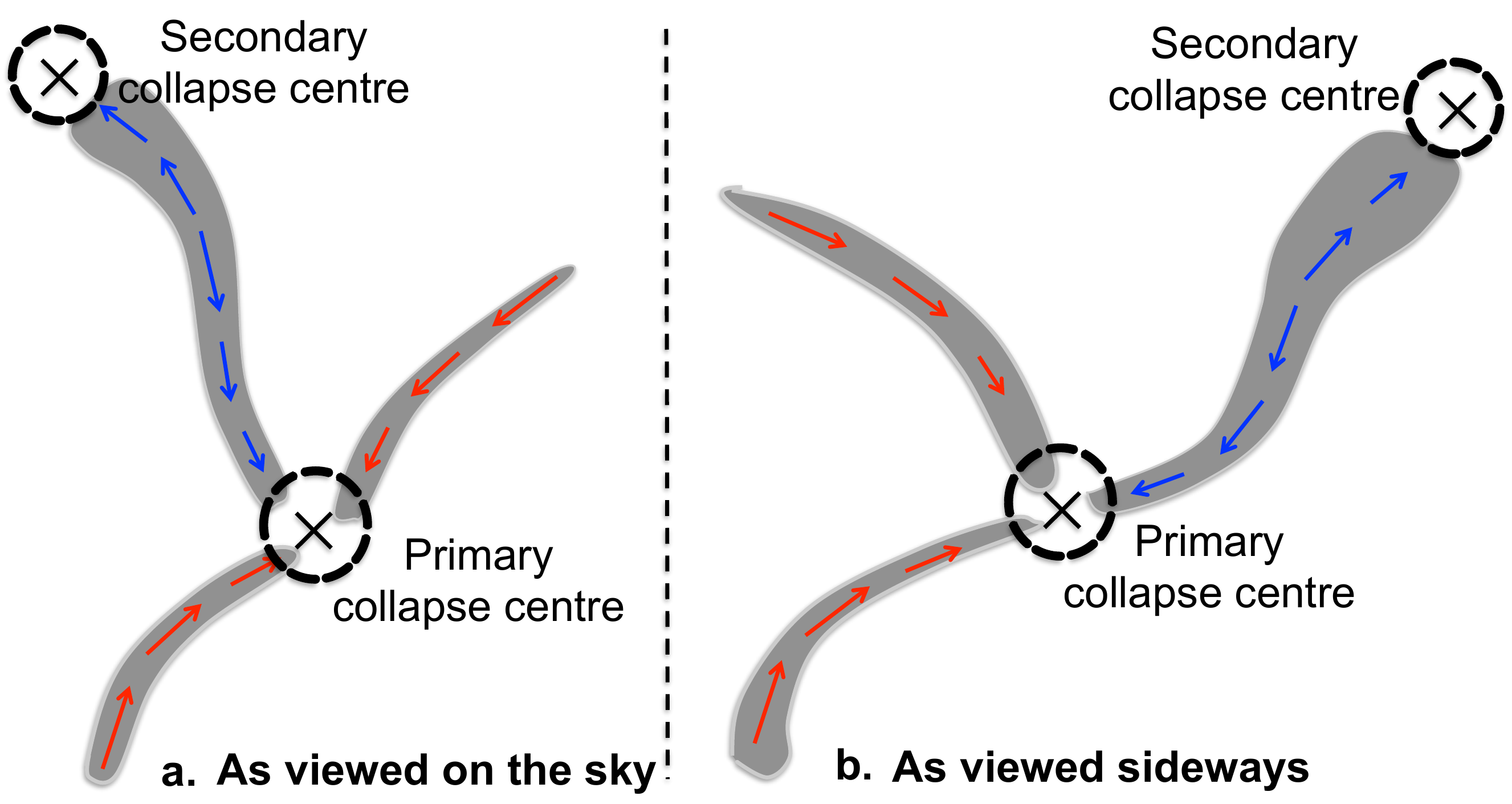} 
          \caption{Schematic representation of the the SDC13 velocity field (arrows) (a) as viewed on the plane of the sky; (b) as viewed sideways with the observer on the left-hand-side of the plot.  The two centres of collapse are symbolised by black dashed circles and a cross at their centres.}
      \label{infpic}
      \end{figure}

In the context of longitudinal collapse the observed velocity gradients are an indication of the time evolution of each filament as the gradient becomes steeper during collapse. We can therefore potentially constrain the age of a filament by measuring its velocity gradient.  \cite{pon2012} showed that the collapse time of free-falling filaments is $\tau_{1D} = \sqrt{2/3}\,A\,\tau_{3D}$ where $\tau_{3D}$ is the spherical free fall time at the same volume density, and $A$ is the aspect ratio of the filament. This equation shows that the collapsing time of a filament can be significantly longer than the standard $\tau_{3D}$. One can compute the time evolution of the filament velocity gradient as a function of the semi-major axis $Z$, and this for a given initial density $n_0$, initial aspect ratio $A_0$, and initial semi-major axis $Z_0$, and then compare these to the SDC13 filament observed values (see Appendix A for more details). Figure~\ref{model} shows that the velocity gradients of modelled filaments  are consistent with the observed ones after 1 to 4~Myr of evolution, depending on the projection angles.  Similar collapsing timescales have recently been reported for the massive-star forming filament NGC6334 \citep{zernickel2013}. Also, the fact Fi-NEn seems dynamically younger is consistent with the idea that it is collapsing towards a secondary potential well centre around position 31. Despite the limitations of this comparison (SDC13 is more of a hub-filament system rather than a single filament - cf Fig.~\ref{infpic}) these timescales are consistent with the presence of Class I protostellar sources in the SDC13 filaments.


\subsection{Gravity-driven turbulence and core formation}

 Figure~\ref{profiles}(right) shows the total velocity dispersion (including thermal and non-thermal motions - cf Arzoumanian et al. 2013) along each filaments. We can see a strong correlation between this quantity and the dense gas velocity, the velocity dispersion increasing towards each of the two centres of collapse (positions 8 and 31). During the collapse of the filaments the gravitational energy of the gas is converted into kinetic energy \citep[e.g.][]{peretto2007,vazquezsemadeni2007}. This has been proposed to be the main process by which radially collapsing filaments manage to keep a constant width \citep{arzoumanian2013}. The velocity dispersion resulting from gravitational energy conversion is then a function of the infall velocity and the gas density \citep{klessen2010}. The global dynamical evolution of a supercritical filament initially at rest, with uniform density, can be described in two stages \citep[e.g.][]{peretto2007}. In the first stage, the gas at the filament ends is accelerated more efficiently, and the filament develops a linear velocity gradient increasing from centre to ends. In the second stage, as the matter is accumulated at the centre, the velocity gradient starts to reverse due to the acceleration close to the central mass becoming  larger than  at the filament ends. It is clear that in the second stage, both the infall velocity and the velocity dispersion will increase faster at the centre than at the filament ends. However, in the first stage we might expect the infall velocity dispersion and the infall velocity to be larger at the filament ends. In the case of SDC13, the infall velocity is larger at the filament ends, but the velocity dispersion is larger at the centre. This could be explained by the fact that the SDC13 filaments are in an intermediate  evolutionary stage, and/or that  the initial density profile of SDC13 is not uniform.  In fact, SDC13 as a whole has a more complex geometry and density structure than the ones of a single filament, it resembles more of a hub-filament system \citep{myers2009}. In such configuration, the potential well is dominated by the central mass of the hub.  The resulting hub density profile might then favour an early increase of the velocity dispersion during the hub collapse. However, this is speculation and remains to be tested. 
 
The MM1 and MM2 sources, the most massive ones in SDC13, are located near the centre of  the system. Within the context of the free-fall collapse scenario these sources must have formed within a tenth of a pc from their current position, suggesting that MM1 and MM2 have basically formed in situ, near the converging point of the filaments. This location is privileged as source mass growth is concerned since flows of dense gas are running towards it, both bringing more material to be accreted, and locally increasing the velocity dispersion. We can quantify what impact both processes have on core formation. The mass infall rate due to material running through the filaments can be estimated following $\dot{M}_{inf}=\pi (W/2)^2\rho_{fil}V_{inf}$ where $W$ is the filament width, $\rho_{fil}$ the filament density, and $V_{inf}$ the infall velocity. Using the average values given in Table 3, along with an infall velocity $V_{inf}\simeq0.2$~km/s at the core locations, we find that $\dot{M}_{inf}\simeq2.5\times10^{-5}$~M$_{\odot}$/yr. This means that over a million years each filament would bring $\sim25~M_{\odot}$ to the SDC13 centre. This is probably an upper limit since the gas did not infall for a million years with its current infall velocity. Now, if we consider that the gravity-driven turbulence acts as a support against gravity then the effective Jeans mass becomes ${\rm M_{J}^{eff}}=0.9~{\rm M_{\odot}} (\sigma_{tot}/0.2~{\rm km\,s^{-1}})^3$.  With a gas velocity dispersion $\sigma_{tot}\simeq 0.8$~km/s towards the centre of the system, the effective Jeans mass goes up to $\rm{M_{J}^{eff}}\simeq 60$~M$_{\odot}$. This is close to the observed masses for MM1 and MM2, and suggests that filament longitudinal collapse might lead to the formation of super-Jeans cores more as a result of enhanced turbulent support than as a result of large-scale accretion. 


\section{Conclusion}

The presence of protostellar sources and organised velocity gradients along the SDC13 filaments suggest that local and large-scale  longitudinal collapse are simultaneously taking place. As a result of the high aspect ratios of the SDC13 filaments, the timescale of the former process is much shorter than that of the latter. Here we propose that the large-scale longitudinal collapse mostly contributes to the increase of the turbulent support towards the centre of collapse where starless cores with masses order of magnitude larger than the thermal Jeans mass can form. This study therefore suggests that super-Jeans prestellar cores may form at the centre of collapsing clouds. Despite this increase of support at the centre of SDC13,  the MM1 and MM2 masses are still much lower than the mass of the O-type star-forming core sitting at the centre of SDC335 \citep{peretto2013}. We speculate that the major difference resides in the mass and morphology of the surrounding gas reservoirs surrounding these sources. Dense, collapsing, spherical clouds of cold gas are more efficient in concentrating matter at their centres in a short amount of time. The study of a larger sample of massive star-forming clouds presenting all sort of morphologies is required to test this assertion.


\begin{table*}
\caption{Observed properties of SDC13 filaments}             
\label{table:1}      
\centering                          
\begin{tabular}{c c c c c }        
\hline\hline                 
Name & Sizes & PA& R$_{eff}$& $F^{int}_{1.2mm}$ \\    
& ($\arcsec\times\arcsec$)& (\degr)&(\arcsec)& (Jy) \\
\hline                        
 Fil-NEn   & $104\times20$ & +38 &  30.6 & $0.259\pm0.032$  \\      
  Fil-NEs   & $95\times19$ & +19 &  29.5 & $0.213\pm0.030$  \\ 
 Fil-NW   &  $152\times16$    & -31& 29.6& $0.198\pm0.031$ \\
 Fil-SE  &  $107\times22$    &  -56& 32.8 & $0.219\pm0.032$ \\
\hline                                   
\end{tabular}
\tablefoot{(col.\,1): Filament name; (col.\,2): Non-deconvolved filament length and filament width, angular size uncertainties are $\pm1\arcsec$; (col.\,3): Filament position angle; (col.\,4): Filament geometric radius; (col.\,5): 1.2mm dust continuum flux.  }
\end{table*}

\begin{table*}
\caption{Observed properties of SDC13 cores}
\label{table:1}
\centering
\begin{tabular}{c c c c c c c c c c c}
\hline\hline
Name & RA & Dec & $F^{pk}_{1.2mm}$& $F^{int}_{1.2mm}$ & Maj. & Min.& PA & R$_{eff}$&  Protostellar?& Fragmented? \\
& (J2000) & (J2000) &(mJy/beam) &(mJy)  & ($\arcsec$) & ($\arcsec$)& (\degr)&(\arcsec)&  &  \\
\hline
 MM1 & 18:14:30.86 & -17:33:20.4 &  51.8 &  $94.5 \pm34.3$ & 28.5 & 11.6 &  +27. & 15.7 & yes &  --     \\
 MM2 & 18:14:28.53 & -17:33:30.9 &  48.3 &  $73.2 \pm23.3$ & 14.0 & 11.9 &  +21 & 12.8 &  no &  no    \\
 MM3 & 18:14:35.53 & -17:30:53.4 &  34.6 &  $22.3 \pm16.5$ & 11.3 &  6.4 &  +26 &  7.7 &  no &  no    \\
 MM4 & 18:14:30.63 & -17:33:59.0 &  31.2 &  $23.0 \pm13.8$ & 10.5 &  6.8 & -49 &  8.4 & yes &  --    \\
 MM5 & 18:14:33.89 & -17:31:14.4 &  28.6 &  $10.3 \pm8.4$ &  8.4 &  3.8 &  +39 &  5.6 & yes &   no   \\
 MM6 & 18:14:35.30 & -17:30:36.0 &  28.4 &   $8.8 \pm 7.7$ &  6.5 &  3.7 & -45 &  5.2 &  no & no \\
 MM7 & 18:14:32.96 & -17:34:19.9 &  26.3 &  $26.0\pm17.6$ & 15.6 &  7.2 & -60 &  9.9 & yes & no     \\
 MM8 & 18:14:29.93 & -17:33:48.4 &  20.4 &  -- &  -- &  --&   -- &  -- &  no & -- \\
 MM9 & 18:14:32.73 & -17:31:39.0 &  19.9 &  $9.8 \pm7.0$ &  7.7 &  5.6 &  +43 &  6.6 &  no &   no   \\
MM10 & 18:14:32.26 & -17:32:24.5 &  18.8 &   $31.0 \pm17.3$ & 18.7 & 10.5 &  +27 & 13.0 &  no & yes \\
MM11 & 18:14:27.37 & -17:32:42.0 &  17.9 &  $5.7 \pm4.0$ &  7.00 &  3.9 & -32 &  5.2 &  no &   no   \\
MM12 & 18:14:34.59 & -17:30:21.9 &  17.9 &  -- &  -- &  -- &   -- &  -- &  no &    --  \\
MM13 & 18:14:25.72 & -17:32:07.0 &  17.8 &  $6.5 \pm4.2$ &  6.1 &  4.8 & -44 &  5.6 &  no &   no   \\
MM14 & 18:14:34.83 & -17:34:41.0 &  17.7 &  $14.0 \pm11.0$ & 10.5 &  8.2 &  -9 &  8.6 &  no &  --    \\
MM15 & 18:14:26.67 & -17:32:27.9 &  16.2 &   -- &  -- &  -- &   -- &  -- &  no &  no \\
MM16 & 18:14:24.56 & -17:31:42.4 &  16.1 &  $8.5 \pm5.6$ & 12.2 &  4.1 & -35 &  6.8 &  no & no \\
MM17 & 18:14:24.79 & -17:31:56.4 &  14.6 &  -- &  -- &  -- &   -- &  -- &  no & no \\
MM18 & 18:14:29.46 & -17:33:13.4 &  13.8 &   -- &  -- &  -- &   -- &  -- & yes & -- \\
\hline
\end{tabular}
\tablefoot{(col.\,1): Source name' (col.\,2 \& 3): Source coordinates; (col.\,4): 1.2mm dust continuum peak flux density; (col.\,5): Source 1.2mm dust continuum integrated flux density. A dash indicates that the source is too weak/blended to extract any robust physical parameter; (col.\,6): Source major axis, angular size uncertainties are $\pm1\arcsec$; (col.\,7) Source minor axis; (col.\,8): Source position angle; (col.\,9): Source geometric radius; (col.\,9): Tag informing  on the protostellar/starless nature of the Source; (col.\,10): Tag informing if a source is sub-fragmented on the 8\micron\  extinction map. A dash indicates that the source is not seen in extinction. }
\end{table*}

\begin{table*}
\caption{Physical properties of SDC13 filaments}             
\label{table:1}      
\centering                          
\begin{tabular}{c c c c c c c}        
\hline\hline                 
Name & Sizes$_{dec}$ & Aspect ratio& Mass &  Density & M$_{line}$ &$\nabla V$ \\    
& ($pc\times pc$)& &(M$_{\odot}$)&($\times10^4$~cm$^{-3}$)& (M$_{\odot}$\, pc$^{-1}$) & (km\,s$^{-1}$\,pc$^{-1}$)\\
\hline                        
 Fil-NEn   & $1.8\times0.3$ & 6 &$256 \pm 32$ & $3.5\pm0.4$ &  $142\pm18$ & $0.22\pm0.07$\\  
 Fil-NEs   & $1.6\times0.3$ & 5 &$211 \pm 30$ & $3.2\pm0.5$ &  $117\pm19$ &$0.63\pm0.19$ \\        
 Fil-NW   &  $2.6\times0.2$& 13 &$195 \pm 31$& $4.1\pm0.7 $ & $75\pm12$ & $0.36\pm0.11$\\
 Fil-SE  &  $1.9\times0.3$   & 6 & $ 216 \pm 32$ & $2.8\pm0.4 $ & $114\pm17$ & $0.62\pm0.14$ \\
\hline                                   
\end{tabular}
\tablefoot{(col.\,1): Filament name; (col.\,2): Deconvolved filament sizes; (col.\,3): Filament aspect ratios. Note that given the dynamics of the SDC13 filaments, i.e. all collapsing towards one of their end, the values given in this column correspond to half of the aspect ratios used for the modelling in Sect.~4.3. (co.\,4): Filament mass; (col.\,5): Filament volume density assuming cylindrical geometry; (col.\,6): Filament mass-per-unit-length; (col.\,7): Filament velocity gradients measured up to a distance of 1.5~pc for all filaments (leaving out the flat part of the velocity profile, see Fig.~\ref{profiles}). }
\end{table*}

\begin{table*}
\caption{Physical properties of SDC13 cores}
\label{table:1}
\centering
\begin{tabular}{c c c c}
\hline\hline
Name & R$_{eff}^{dec}$ & Mass & Density \\
& (pc) & (M$_{\odot}$) &($\times10^4$~cm$^{-3}$)   \\
\hline
 MM1 & 0.26 & $74.8\pm27.1$ &  $1.8\pm0.6$    \\
 MM2 & 0.21 & $81.1\pm25.8$  &   $3.6\pm1.1$  \\
 MM3 & 0.10 & $24.7\pm18.3$ & $10.2\pm7.6$\\
 MM4 & 0.11 & $18.1\pm10.9$ & $5.6\pm3.4$ \\
 MM5 & $<0.05$& $8.2\pm6.6$ & $>27.2\pm21.9$\\
 MM6 & $< 0.05$ & $9.7\pm8.5$ &$>32.1\pm28.2$ \\
 MM7 & 0.15 & $20.2\pm13.9$ & $2.5\pm1.7$  \\
 MM8 & -- & -- & -- \\
 MM9 & 0.06 &  $10.9\pm7.7$ & $20.9\pm14.8$ \\
MM10 & 0.21 &$34.3\pm19.2$ & $1.5\pm0.9$ \\
MM11 & $<0.05$ & $6.3\pm4.4$ & $>20.9\pm14.6$  \\
MM12 & -- & -- &--  \\
MM13 & $<0.05$ &$7.3\pm4.7$ &  $>24.2\pm15.6$  \\
MM14 & 0.12 & $15.6\pm12.2$ &   $3.8\pm2.9$\\
MM15 & -- & -- & --\\
MM16 & 0.07 & $9.4\pm6.1$ & $11.3\pm7.4$ \\
MM17 & -- & -- & --  \\
MM18 & -- & -- & --  \\
\hline
\end{tabular}
\tablefoot{(col.\,1): Source name; (col.\,2): Source deconvolved geometric radius; (col.\,3): Source mass; (col.\,4): Source volume density estimated assuming spherical geometry.}
\end{table*}

\begin{table*}
\caption{N$_2$H$^+$(1-0) spectra properties}
\label{table:1}
\centering
\begin{tabular}{c c c c c c c}
\hline\hline
Id & RA & Dec & Velocity & FWHM & $\sigma_{NT}$ & $\sigma_{tot}$\\
& (J2000) & (J2000) &(km/s) &(km/s)  & km/s & km/s \\
\hline
 P1 & 18:14:35.29 & -17:34:41.9 &  37.83 &  0.87 & 0.37& 0.42\\
 P2 & 18:14:34.50 & -17:34:33.7 &  37.89 &  0.95  & 0.40 & 0.45\\
 P3 & 18:14:33.77 & -17:34:25.8 &  37.88 &  1.32 & 0.56 & 0.60\\
 P4 & 18:14:32.95 & -17:34:18.0 &  37.68 &  1.58 & 0.67 & 0.70\\
 P5 & 18:14:32.16 & -17:34:10.3 &  37.58 &  1.60 & 0.68 & 0.71\\
 P6 & 18:14:31.36 & -17:34:02.8 &  37.38 &  1.57 & 0.67 & 0.70\\
 P7 & 18:14:30.54 & -17:33:58.1 &  37.25 &  1.76 & 0.75 & 0.78\\
 P8 & 18:14:29.69 & -17:33:49.4 &  37.09 &  1.87 & 0.79 & 0.82\\
 P9 & 18:14:28.53 & -17:33:38.0 &  37.19 &  1.92 & 0.81& 0.84 \\
P10 & 18:14:28.11 & -17:33:25.8 &  37.24 &  1.97 & 0.84& 0.86\\
P11 & 18:14:27.84 & -17:33:12.7 &  37.48 &  1.50 & 0.64 & 0.67\\
P12 & 18:14:27.50 & -17:32:59.9 &  37.41 &  1.22 & 0.52 & 0.56\\
P13 & 18:14:27.29 & -17:32:46.1 &  37.49 &  1.03 & 0.43 & 0.48\\
P14 & 18:14:26.71 & -17:32:35.2 &  37.60 &  1.03 & 0.43 & 0.48\\
P15 & 18:14:25.34 & -17:32:15.0 &  37.39 &  0.68 & 0.28 & 0.35\\
P16 & 18:14:30.74 & -17:33:38.2 &  36.83 &  1.72 & 0.73 & 0.76\\
P17 & 18:14:29.58 & -17:33:32.2 &  36.96 &  1.86 & 0.79 & 0.82\\
P18 & 18:14:30.83 & -17:33:18.7 &  36.82 &  1.85 & 0.78 & 0.81\\
P19 & 18:14:31.09 & -17:33:05.9 &  36.59 &  1.52 & 0.64 & 0.67\\
P20 & 18:14:31.36 & -17:32:52.8 &  36.48 &  1.35 & 0.57 & 0.61\\
P21 & 18:14:31.67 & -17:32:39.7 &  36.33 &  0.88 & 0.37 & 0.42\\
P22 & 18:14:32.01 & -17:32:26.9 &  36.19 &  1.06 & 0.45 & 0.49\\
P23 & 18:14:32.17 & -17:32:13.1 &  36.21 &  1.01 & 0.43 & 0.48\\
P24 & 18:14:32.17 & -17:31:59.6 &  36.17 &  0.87 & 0.37 & 0.42 \\
P25 & 18:14:32.17 & -17:31:45.7 &  36.18 &  0.66 & 0.27 & 0.34\\
P26 & 18:14:32.38 & -17:31:32.6 &  36.19 &  0.66 & 0.27 & 0.34\\
P27 & 18:14:32.93 & -17:31:20.6 &  36.34 &  0.79 & 0.33 & 0.39\\
P28 & 18:14:33.61 & -17:31:10.8 &  36.37 &  1.01 & 0.43 & 0.48\\
P29 & 18:14:34.47 & -17:31:04.4 &  36.44 &  1.29 & 0.55 & 0.59 \\
P30 & 18:14:35.23 & -17:30:56.2 &  36.44 &  1.38 & 0.58 & 0.62 \\
P31 & 18:14:35.65 & -17:30:43.1 &  36.47 &  1.44 & 0.61 & 0.64 \\
\hline
\end{tabular}
\tablefoot{The uncertainties on all velocity measurements are $\pm0.05$~km/s. (col.\,1): Spectrum position number (as displayed in Figs.~\ref{spectra} and online \ref{allspec}); (col.\,2 \& 3): Position coordinates; (col.\,4): Spectrum best-fit velocity; (col.\,5): Spectrum best-fit line width; (col.\,6): Spectrum non-thermal velocity dispersion component; (col.\,7): Spectrum total (thermal+non-thermal) velocity dispersion. We assumed a gas temperature of 12~K for these calculations. }
\end{table*}

\begin{acknowledgements}
We thank the anonymous referee whose report helped improving the quality of this paper. We would like to thank Alvaro Hacar for helping with the MAMBO data reduction. And finally, we want to thank the IRAM 30m staff for their support during the observing runs.

\end{acknowledgements}

\bibliographystyle{aa}
\bibliography{references}

\begin{appendix}

\section{Homologous free-fall collapse of filaments}

The equation of motion for a filament edge in homologous free-fall collapse is given by \citep{pon2012}:

\begin{equation}
\frac{d^2Z(t)}{dt^2}=\frac{GM}{Z(t)^2}
\end{equation}
where $Z(t)$ is the semi-major axis of the filament, and M its mass. Multiplying each side of Eq.~(A.1) by $\frac{dZ(t)}{dt}$ and integrating over $t$, we obtain the following equation:

\begin{equation}
\frac{dZ(t)}{dt}=V(t)=-\sqrt{2GM[Z(t)^{-1}+\alpha]} 
\end{equation}
where $\alpha$ is the constant of integration. Now, if we consider a filament initially at rest, we have $V(t=0)=0$, and therefore $\alpha= -Z_0^{-1}$, where $Z_0=Z(t=0)$ is the initial semi-major axis of the filament. We can solve for $Z(t)$ by setting $Z(t)/Z_0=\cos^2\beta$ in Eq.~(A.2) and integrating over $t$ again. We then obtain the following equation:

\begin{equation}
\beta+\frac{1}{2}\sin2\beta = t\sqrt{2GM/Z_0^3}
\end{equation}
or when replacing the mass by $M=2\pi R^2Z_0\rho_0$

\begin{equation}
\beta+\frac{1}{2}\sin2\beta = \frac{t}{A_0}\sqrt{4\pi G\rho_0}
\end{equation}
where $A_0$ is the initial aspect ratio of the filament and $\rho_0$ its initial density.  Note that the filament free-fall times $\tau_{1D}$  is calculated for $Z(t)/Z_0=0$, which means when $\beta=\pi/2$. We therefore obtain:

\begin{equation}
\tau_{1D}=A_0\sqrt{\frac{\pi}{16G\rho_0}}=\sqrt{2/3}A_0\tau_{3D}
\end{equation}
where $\tau_{3D}$ is the spherical free-fall time at the same density. Finally, given that the velocity gradient is linear during the homologous collapse of a filament, we can compute the velocity gradient $\nabla V=V(t)/Z(t)$.

 In order to compute the different values plotted in Fig.~\ref{model} we sampled $\beta$ between 0 and $\pi/2$ and used the relevant equations presented in this appendix. We further assumed that the radius of the filaments stay the same during the collapse. For the purpose of the calculations we took $R=0.15$~pc. Figure~\ref{model} shows the evolution of seven filaments (solid lines) having all the same initial density $n_0=4\times10^4$~cm$^{-3}$ but a different initial aspect ratio\footnote{Note that since we fixed the filament radius, $A_0$ and $Z_0$ are not independent parameters. Also, the aspect ratios given in Table 3 correspond to half of this $A_0$ parameter.}.

Nearly all quantities presented in this paper are affected by projection effect. We can only correct for it if we know the projection angle $\theta$ of the filaments with respect to the line of sight. Unfortunately, it is impossible to know the exact value of $\theta$ for a given filament. However, for randomly sampled filament orientations one have $<\theta>=67\degr$. So for this letter, we considered two cases, the case where $\theta=45\degr$ and the case where $\theta=67\degr$. The corrected values for the velocity, semi-major axis, and velocity gradients are: $V_{corr}= V_{obs}/\cos(\theta)$,  $L_{corr}=L_{obs}/\sin(\theta)$, and $\nabla V_{corr}=\nabla V_{obs} \tan(\theta)$. For $\theta=45\degr$ these corrections meant hat the observed velocity gradients remain unaffected, but the semi-major axis becomes larger by a factor 1.4. For $\theta=67\degr$, the velocity gradient increases by a factor 2.4, and the semi-major axis by a factor 1.1. The points in Fig.~\ref{model} corresponding to the four SDC13 filaments have been corrected by such factors. Note that much larger angles would lead to very large velocity gradients which have never been observed on parsec-scales, while much smaller angles would imply very large initial aspect ratios and semi-major axes. Even though we cannot rule out such configurations, they are unlikely.

\section{Protostellar classes of SDC13 sources}
 
Infrared spectral indexes have been used for more than 20 years to
classify YSOs according to their evolutionary stages. The idea is
that the peak of a YSO spectral energy distribution (SED) evolves from submillimeter to infrared during
the first few million years of its life. Calculating then the slope, i.e. the
spectral index, of the SED at a critical frequency range allows the quantification of
this evolution. YSOs classes  are traditionally  defined \citep[e.g.][]{lada1984,wilking1989} between 2\micron\  and 10\micron\ by calculating the following quantity  
 $\alpha_{[2-10]} = \frac{d(\log(\lambda
  S_{\lambda}))}{d(\log(\lambda))}$, Class I
sources having  $\alpha_{[2-10]} > 0$, Class II sources $0>
\alpha_{[2-10]} > -1.6$, and Class III sources $\alpha_{[2-10]} <
-1.6$.  

\begin{figure}
   \centering   
    \includegraphics[width=9.cm]{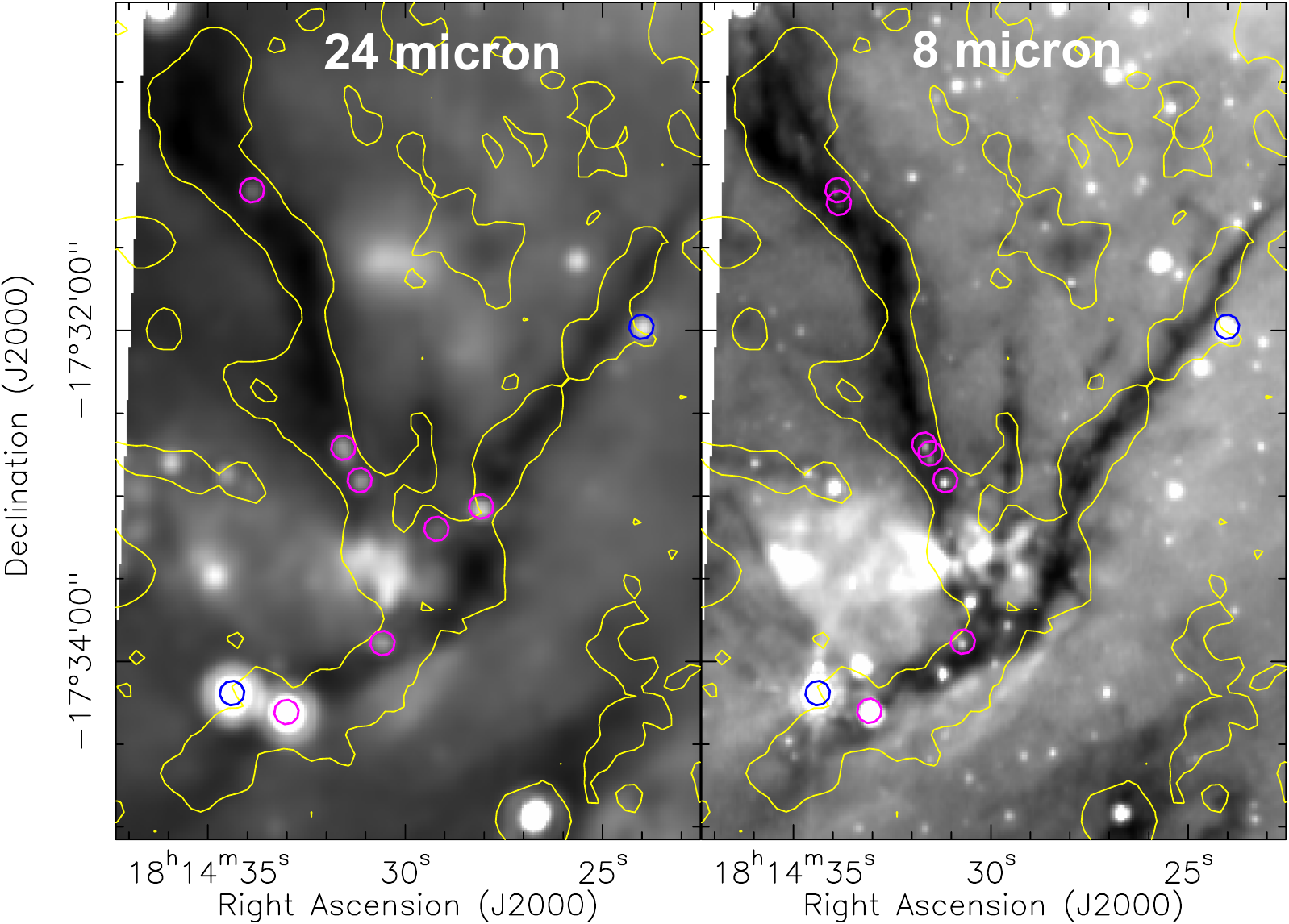}
      \caption{(left): Spitzer 24\micron\ image of SDC13 (greyscale) on which we over plotted the MAMBO 3~mJy/beam contour (yellow). On this image we also report the positions of the seven protostellar sources identified alone the SDC13 filaments (purple circles), along with the detection of two extra sources located on the edge of the outer edge of the contour (blue circles). (right): 8\micron\ image of SDC13 (greyscale). The contour is the same as in the left panel. The purple circles mark the positions of the nine 8\micron\ sources identified towards the identified 24\micron\ sources. The meaning of the symbol's colour is the same as in the left panel.}
         \label{24mic}
   \end{figure}

In the context of this study we focussed on sources identified at  24\micron. We only considered sources which are spatially coincident with the SDC13 filaments (purple circles on Fig.~\ref{24mic}), limiting potential background/foreground contamination. To extract sources we used the Starfinder algorithm \citep{diolaiti2000} and found seven 24\micron\ sources, five of which have 8\micron\ counterparts\footnote{Note that two 24\micron\ sources split up into two 8\micron\ sources} (see Fig.~\ref{24mic}). We then calculated their spectral index as estimated between  8\micron\ and 24\micron (fluxes of multiple 8\micron\ sources have been combined, and for those with no 8\micron\ counterpart we used the 8\micron\ detection limit of 2~mJy). One advantage provided by these two wavelengths is that they are similarly affected by extinction \citep{flaherty2007,chapman2009} which implies that estimated values of $\alpha_{[8-24]}$ are not much affected by extinction.  As a result, we find three sources with $-0.4<\alpha_{[8-24]}<0$, and another four with $0<\alpha_{[8-24]}<2.2$. Note that we also report the detection of two additional protostellar sources located on the outer edge of the filament (blue circles in Fig.~\ref{24mic}). Their spectral indexes are $\sim -1.5$, indicating that they might well be background/foreground sources. Note that we do not have any detection towards MM1 at 24\micron. This is due to the fact that the 24\micron\ emission is rather diffuse towards this source, and Starfinder failed to find any compact sources, a requirement for protostar detection. 

\begin{figure}
   \centering   
    \includegraphics[width=7.cm]{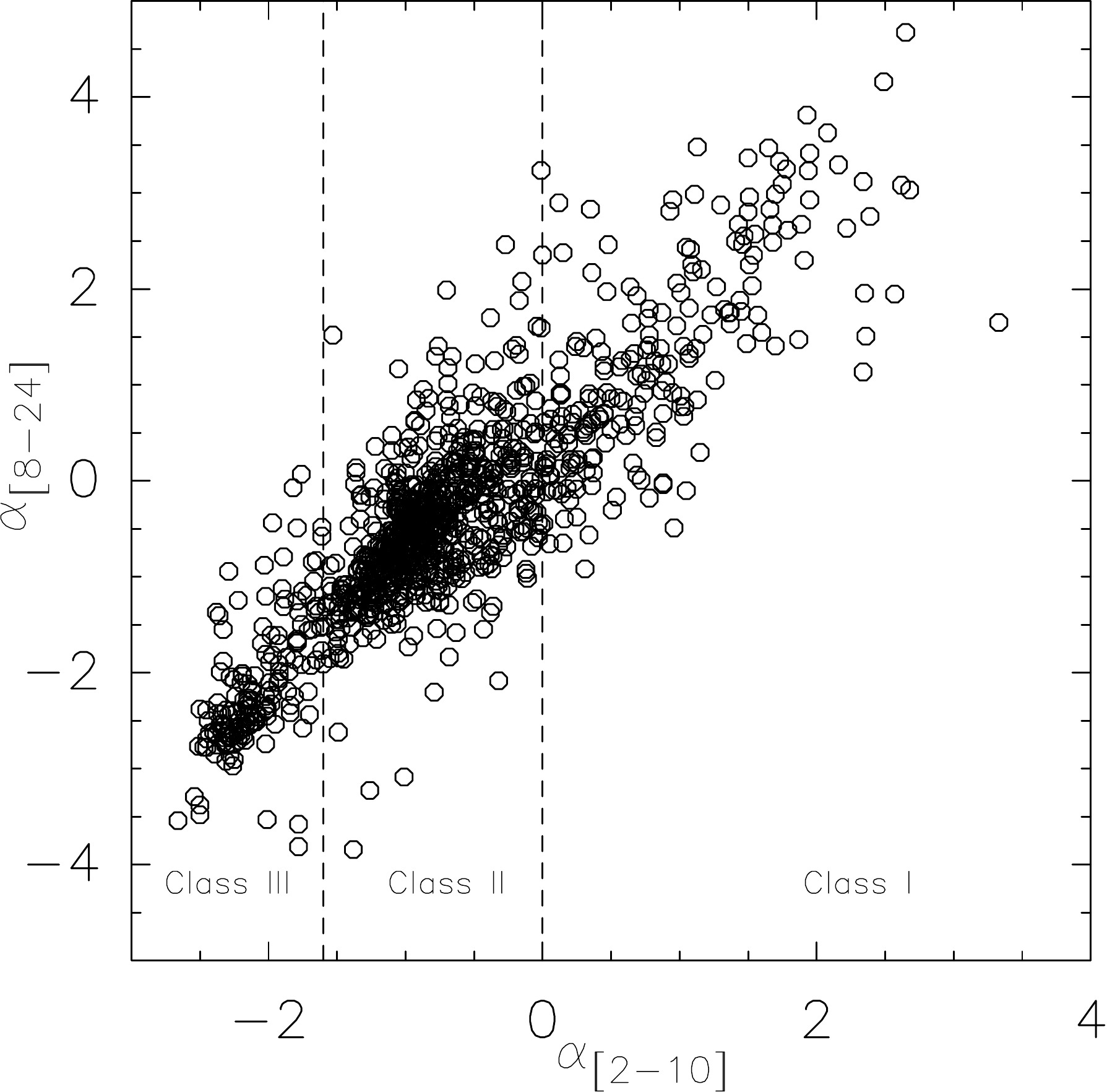}
      \caption{Mid-infrared spectral indexes for all $\sim1000$ YSOs from the C2D catalogue \citep{evans2009}. For each source the spectral index has been calculated for two wavelength ranges: from 2\micron to 10\micron, and from 8 \micron to 24\micron. }
         \label{index_comp}
   \end{figure}

We then used  the YSO C2D database of  Serpens, Ophiucchi, and Perseus
\citep{evans2009} in order  to compare the $\alpha_{[2-10]}$ spectral index
with $\alpha_{[8-24]}$. The results of this comparison are displayed in Fig.~\ref{index_comp}.
Although there is some dispersion and not a one-to-one correlation,
there is still a linear correlation between the two indexes. Despite the dispersion in $\alpha_{[8-24]}$ this plot strongly suggests  that the seven protostellar sources identified within the SDC13 filaments have mid-infrared spectral indexes consistent with Class I and/or young Class II objects.

\begin{figure}
   \centering   
    \includegraphics[width=7.cm]{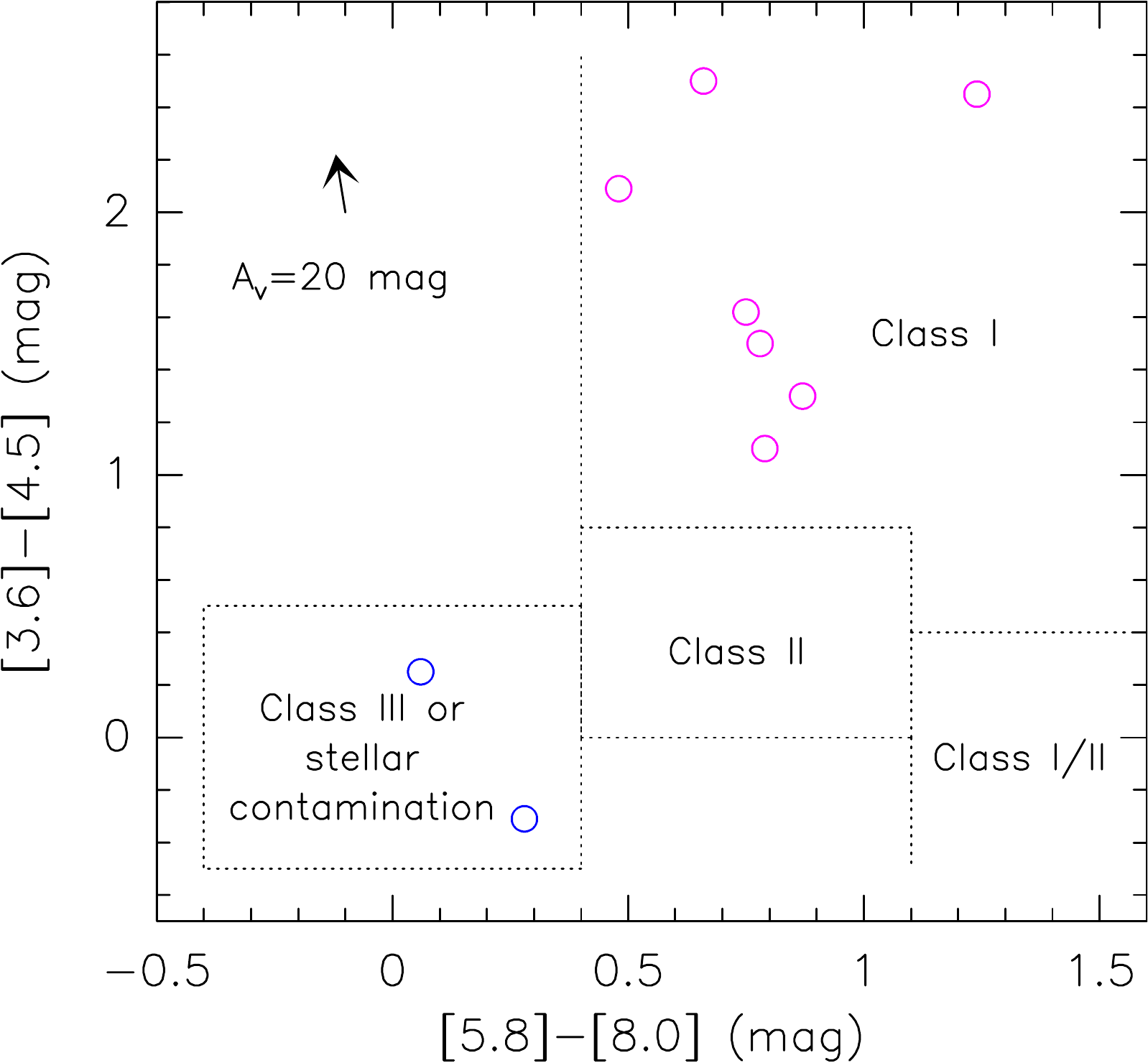}
      \caption{Mid-infrared [3.6]-[4.5] vs [5.8]-[8.0] colour-colour diagram for the nine sources extracted at 8\micron. The colour coding is the same as in Fig.~\ref{24mic}. The arrow indicates an extinction of $A_v=10$. The different regions of the plot corresponding to the different protostellar classes are also shown.  }
         \label{colcol}
   \end{figure}

We further analysed the properties of the nine 8\micron\ sources (see Fig/~\ref{24mic}) by constructing a colour-colour diagram using all IRAC {\it Spitzer} bands between 3.6\micron\ and 8\micron. Depending on their colours, sources of different classes are expected to sit in different regions of such diagram \citep{allen2004,megeath2004}. On Fig.~\ref{colcol} we can see that seven sources sit in the Class I region of the diagram, while the two remaining sources  are clearly sitting in the Class III/stellar contamination region. The latter are in fact the two sources located on the edge of the filament (i.e the blue circles in Fig.~\ref{24mic}), already identified as likely foreground/background sources from our spectral index analysis. Note also that extinction can significantly affect the source location in this plot, mostly artificially increasing their [3.6] -[ 4.5] magnitude. However, the extracted  sources are red enough that even for an extinction of $A_v=50$ most identified sources would remain in the Class I region of the diagram, the other ones just moving  to the other side of the Class II/Class I border.  This is consistent with the $\alpha_{[8-24]}$ analysis performed above.

Finally, based on the 24\micron\ flux of protostellar sources and an assumed extinction, one may obtain an estimate of the source bolometric luminosity \citep{dunham2008}. Following \citet{parsons2009} we estimate that, with extinctions in the range of $10<\rm{A_v}<50$ (matching the range of observed dust column density in SDC13) and 24\micron\  fluxes between 10~mJy and 60~mJy for all but one source, the SDC13 protostellar sources have bolometric luminosities in the range of 10~L$_{\odot}$ to few 100~L$_{\odot}$. Such luminosities stand in the high end of the luminosity distribution of nearby low-mass protostars \citep{evans2009}, possibly indicating a slightly shorter lifetime of the SDC13 protostellar sources compared to the standard Class I low-mass protostar lifetime.

\end{appendix}

\Online

\onlfig{
  \begin{figure*}
   \vspace{-0cm}
   \hspace{2.5cm}
   \includegraphics[width=10.5cm,angle=0]{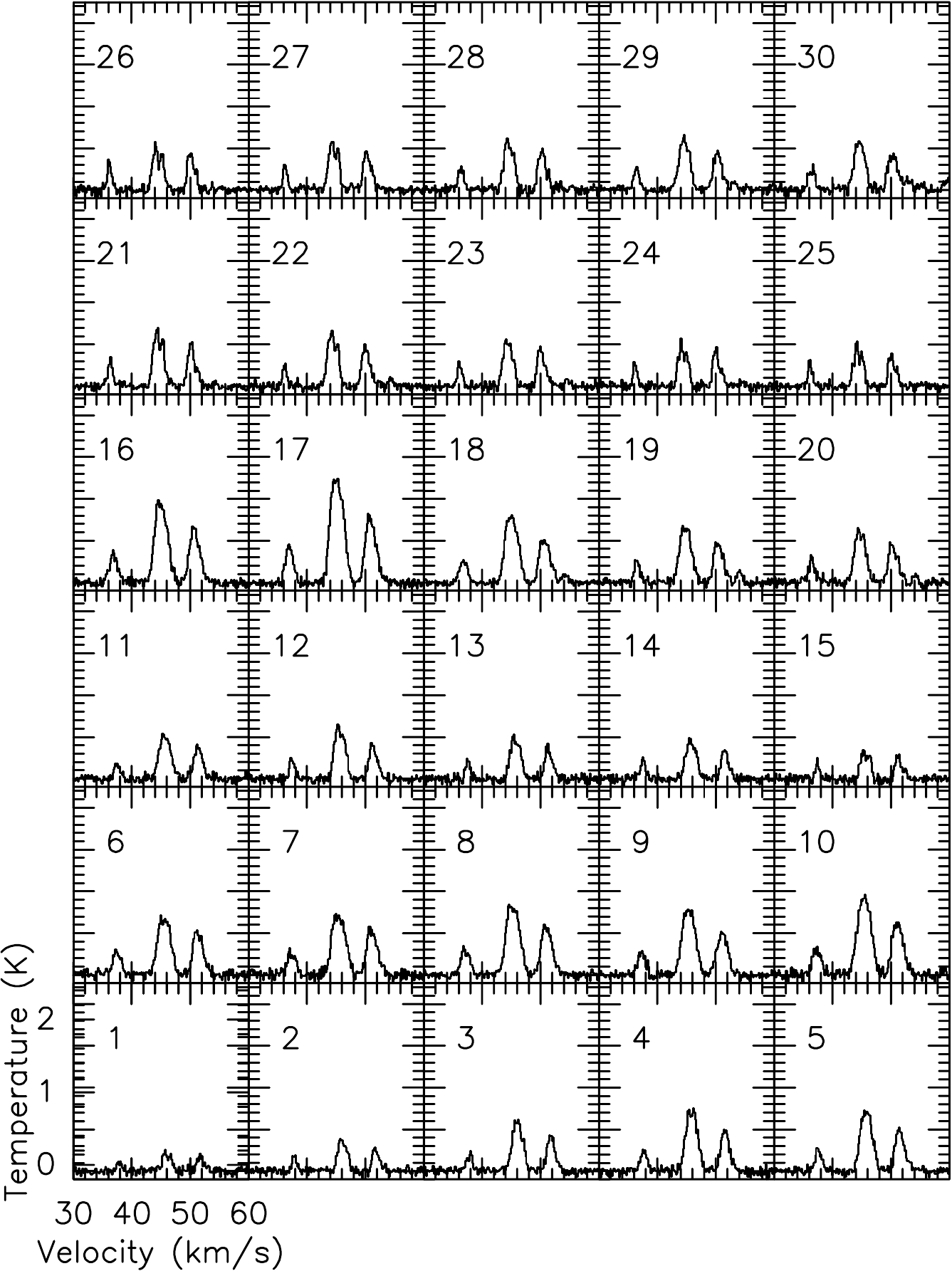} 
   \vspace{-0cm}
      \caption{N$_2$H$^+$(1-0) spectra observed for each position along the SDC13 filaments (see Fig.~2).
              }
         \label{allspec}
   \end{figure*}
}

\end{document}